\documentclass[aps,prd,10pt,notitlepage,nofootinbib,superscriptaddress,showkeys,showpacs]{revtex4-1}
\usepackage{amsmath,amssymb,amsthm,latexsym}
\usepackage[english]{babel}
\usepackage{graphicx,color}
\usepackage{xspace}
\usepackage{graphicx}
\usepackage{pifont,dsfont}
\usepackage{marvosym}
\usepackage{slashed}
\usepackage{subfigure}

\newcommand{\be}{\begin{equation}}
\newcommand{\ee}{\end{equation}}
\newcommand{\bqa}{\begin{eqnarray}}
\newcommand{\eqa}{\end{eqnarray}}
\newcommand{\bea}{\begin{eqnarray}}
\newcommand{\eea}{\end{eqnarray}}

\newcommand{\Tr}{{\rm Tr}}

\newtheorem{theorem}{Theorem}

\newcommand{\cT}{{\cal T}}
\newcommand{\cG}{{\cal G}}
\newcommand{\cB}{{\cal B}}

\newcommand{\cV}{{\cal V}}

\newcommand{\cL}{{\cal L}}

\begin{document}

\title{\Large \bf The Schwinger Dyson equations and the algebra of constraints of random tensor models at all orders}

\author{{\bf Razvan Gurau}}\email{rgurau@perimeterinstitute.ca}
\affiliation{Perimeter Institute for Theoretical Physics, 31 Caroline St. N, ON N2L 2Y5, Waterloo, Canada}

\date{\small\today}

\begin{abstract}
\noindent Random tensor models for a generic complex tensor generalize matrix models in arbitrary dimensions and yield a 
theory of random geometries. They support a $1/N$ expansion dominated by graphs of spherical topology. Their Schwinger Dyson equations, 
generalizing the loop equations of matrix models, translate into constraints 
satisfied by the partition function. The constraints have been shown, in the large $N$ limit, to close a Lie algebra indexed by colored rooted 
$D$-ary trees yielding a first generalization of the Virasoro algebra in arbitrary dimensions. In this paper we complete 
the Schwinger Dyson equations and the associated algebra
at all orders in $1/N$. The full algebra of constraints is indexed by $D$-colored graphs, and the leading order 
$D$-ary tree algebra is a Lie subalgebra of the full constraints algebra.
\end{abstract}

\medskip

\noindent  Pacs numbers: 02.10.Ox, 04.60.Gw, 05.40-a

\maketitle

\section{Introduction}

Random matrix models encode a theory of random Riemann surfaces dual to the ribbon Feynman graphs 
generated by their perturbative expansion \cite{Di Francesco:1993nw}. The amplitudes of the Feynman graphs depend on the size of
the matrix, $N$,
and the perturbative series can be reorganized in powers of $1/N$ \cite{'tHooft:1973jz}. At leading order only planar graphs 
contribute \cite{Brezin:1977sv}.
The planar family is summable and undergoes a phase transition to continuous Riemann surfaces \cite{Kazakov:1985ds,mm}
when the coupling constant approaches some critical value. 
Single and multi-matrix models have been very successful in describing the critical behavior of two-dimensional 
statistical models on random geometries \cite{Kazakov:1986hu, Boulatov:1986sb, Brezin:1989db,Kazakovmulticrit}
and via the KPZ correspondence \cite{Knizhnik:1988ak, david2, DK, Dup} on fixed geometries.
Subleading terms in the $1/N$ expansion can be access through double scaling limits \cite{double,double1,double2}. 

Random matrices are analyzed with many techniques, like gauge fixing to the eigenvalues of the matrix, orthogonal 
polynomials and so on \cite{Di Francesco:1993nw}. 
Of particular interest for this paper are the Schwinger Dyson equations (SDE) (or loop equation) 
\cite{Ambjorn:1990ji,Fukuma:1990jw,Makeenko:1991ry,Dijkgraaf:1990rs} which translate into a
set of Virasoro constraints satisfied by the partition function.

The success of matrix models in describing random two dimensional surfaces inspired their 
generalization in higher dimensions to random tensor models \cite{ambj3dqg,sasa1,mmgravity} (see also 
\cite{sasaa,sasab,sasac,sasad,Oriti:2011jm,Baratin:2011aa,Livine:2011yb}
for more recent developments). The corresponding theory of random higher 
dimensional geometries was initially hoped to give insights into conformal field theory, 
statistical models in random geometry and quantum gravity in three and four dimensions. In spite of these initial 
high hopes, tensor models have for a long time been unsuccessful in providing an analytically controlled
theory of random geometries: until recently all the nice aspects of matrix models could not be generalized
to tensors, as their $1/N$ expansion was missing. 

The situation has changed with the discovery of 
{\it colored} \cite{color,PolyColor,lost} rank $D\ge 3$ random tensor models\footnote{In $D=2$ 
colored matrix models have been studied \cite{difrancesco-rect}, but the colors do not play the 
key role they play in three and more dimensions.  }. 
Their perturbation series supports a $1/N$ expansion \cite{Gur3,GurRiv,Gur4}, indexed by the {\it degree}, 
a positive integer which plays in higher dimensions the role of the genus. We emphasize that the degree is {\it not} a topological 
invariant. Leading order 
graphs, baptized {\it melonic} \cite{Bonzom:2011zz}, triangulate the $D$-dimensional sphere in any dimension \cite{Gur3,GurRiv}. 
They form a summable family as they map to colored rooted D-ary trees \cite{Bonzom:2011zz}. Colored random tensors \cite{coloredreview} 
gave the first analytically accessible theory of random geometries in three and more dimensions and became a rapidly
expanding field \cite{sefu2,Geloun:2011cy,Geloun:2010nw,Bonzom:2011br,Ryan:2011qm,Carrozza:2011jn,caravelli}, 
(see also \cite{rey} for some related developments). In particular the first applications of random tensors to statistical models in random
geometries have been explored \cite{IsingD,EDT,doubletens,Bonzom:2012sz}.

The results obtained for the colored models have been shown to hold in fact for tensor models for a single, complex, 
generic (that is non symmetric) tensor \cite{Bonzom:2012hw}. The colors become a canonical bookkeeping 
device tracking the indices of the tensor. The theory thus obtained is universal \cite{Gurau:2011kk} 
and constitutes the only analytically controlled generalization of matrix models to higher dimensions we have so far.

A posteriori one understands why the discovery of the $1/N$ expansion for tensor models has 
taken so long: in order to have any control on the amplitude of graphs one needs to be able
to track the indices of the tensors (this is achieved by the colors). This is impossible if one 
uses tensors with symmetry properties under permutations of the indices. Tensor models for generic tensors
have been generalized to tensor field theories, \cite{BenGeloun:2011rc,BenGeloun:2012pu,Rivasseau:2011hm} 
in which the quadratic part is not invariant under the $\otimes^D U(N)$ symmetry. The non trivial quadratic part generates a renormalization 
group flow and the $1/N$ expansion is recovered dynamically. We already possess an example of a tensor field theory which is 
asymptotically free \cite{BenGeloun:2012pu}. Due to the universality \cite{Gurau:2011kk} of tensor measures
one is tempted to conjecture that asymptotic freedom will be a feature of all tensor field theories.

In order to study this universal theory of random tensors one must generalize to higher dimensions 
as many of the tools which were so useful in matrix models as possible. This turns out again to be a non trivial problem. For instance, 
although various generalizations of eigenvalues to tensors have been proposed, they lack 
most of the interesting properties of the eigenvalues of matrices (for one the tensor does not diagonalize under
a change of basis). A generalization of the notion of determinant to tensors (formats), the Gel'fand hyperdeterminant \cite{Gelfand}
exist (and interestingly enough it requires that the tensor indices are distinguished), however one does not 
possess an explicit formula for it (except for the (1,1,1) format, in which case one gets just the well known 
Cayley hyperdeterminant). However, $2$ is a rather small value for a supposedly large $N$.

The only set of techniques which can be generalized relatively straightforwardly to tensors are the
SDE's. At leading order the SDE's of tensor models have been derived in \cite{Gurau:2011tj}. As one writes
a SDE for every observable of the model, the equations are indexed by the observables.
At leading order only melonic observables contribute and, as they map to trees, the SDE's and associated
Lie algebra of constraints are indexed by trees. The definition of the algebra relies on a gluing operation
for trees, $\cT_1\star_V\cT_2$, which generalizes the addition of integers (in terms of composition of 
observables for matrix models the addition encodes the gluing of two cycles of lengths $p$ and $q$)
to an ``addition'' of observables indexed by trees. Thus the Virasoro (strictu sensu the positive part of the Witt) 
algebra of constrains of matrix models becomes 
\bea
 \bigl[ L_m,L_n\bigr] = (m-n) L_{m+n} \Rightarrow \bigl[\cL_{\cT_1} ,\cL_{\cT_2} \bigr] = \sum_{V\in \cT_1}
  \cL_{\cT_1 \star_V \cT_2} - \sum_{V\in \cT_2} \cL_{\cT_2 \star_V \cT_1} \; .
\eea

In \cite{Gurau:2011tj} it is proved that the constrains hold at leading order in $N$, 
that is $ \lim_{N \to \infty} N^{-D} \frac{1}{Z} \cL_{\cT}Z = 0$, with $Z$ the partition function of tensor models. 
In this paper we complete the algebra of constrains to all order in $1/N$. This time the observables 
(hence the SDE's and the algebra of constraints) are indexed by $D$-colored graphs $\cB$ with a distinguished 
vertex $\bar v$. The observables can be glued by a $\star_{(v,\bar v')}$ operation and we obtain a Lie algebra of constraints
\bea
  \cL_{ (\cB_1,\bar v_1) } Z = 0 \; , \qquad  \Bigl[  \cL_{(\cB_1,\bar v_1)}   , \cL_{(\cB_2,\bar v_2)}     \Bigr] =
 \sum_{v\in \cB_1 }\cL_{ \bigl(  \cB_1  \star_{ ( v,\bar v_2 ) } \cB_2 , \bar v_1   \bigr) }
  - \sum_{v\in \cB_2} \cL_{   \bigl(  \cB_2  \star_{ ( v,\bar v_1 ) } \cB_1 , \bar v_2   \bigr)       } \; ,
\eea
obeyed at all orders in $N$. When restricted to melonic observables the $\star_{(v,\bar v)}$ composition of observables particularizes 
to the $\star_V$ composition of the associated trees. The melonic observables are closed under this gluing. We thus 
not only dispose of the full set of constraints for random tensor models, but, non trivially, the $D$-ary tree subalgebra 
closes into a Lie subalgebra of the full constraints algebra {\it at all} orders in $1/N$. 

The algebra of constraints is the starting point of the study of the continuum limit \cite{Bonzom:2012hw}
of tensor models. In matrix models the continuum theory, Liouville gravity coupled with various matter fields at a 
conformal point, is identified using the theory of (unitary and non unitary) representations of the Virasoro algebra. The continuum operators, 
which are composite operators in terms of the loop observables, are  
hard to identify \cite{Dijkgraaf:1990rs}. One uses the theory of representations of the Virasoro algebra 
to infer the appropriate correspondence. The same must be done in higher dimensions. We need first to 
study and classify the central extensions and unitary representations of this algebra. Once 
the field content of the continuum theory is identified, we need to translate the continuum observables in terms 
of the observables of tensor models.
This study will provide a landscape of universality classes of continuum theories of random geometries 
accessible using tensor models in arbitrary dimensions.
The algebra of constraints identified in 
\cite{Gurau:2011tj} and completed in this paper at all order constitutes the first step towards 
applying tensor models to the study of conformal field theories and quantum gravity in arbitrary dimensions. 

This paper is organized as follows. In section \ref{sec:model} we recall the generic one tensor models and their
$1/N$ expansion. In section \ref{sec:SDE} we define two graphical operations on arbitrary observables of such models,
derive the SDE's and prove that they close a Lie algebra.

\section{The $1/N$ Expansion of Tensor Models}\label{sec:model}

\subsection{Tensor invariants and action}

Let $H_1, \cdots H_D$ be complex vector spaces of dimensions $N_1 , \cdots N_D$. A covariant tensor $T_{n_1\dots n_D}$ of rank $D$ 
(or a $(N_1-1,\dots N_D-1)$ format) is a collection of $\prod_{i=1}^{D} N_i$ complex numbers supplemented with the requirement 
of covariance under base change. We consider tensors $T$ transforming under the {\it external} tensor product of fundamental 
representations of the unitary group $\otimes_{i=1}^D U(N_i)$, that is each $U(N_i)$ acts independently on its corresponding $H_i$. 
The complex conjugate tensor $\bar T_{  n_1 \dots n_D }$ is then a rank $D$ contravariant tensor. They transform as
\bea
 T'_{a_1\dotsc a_D} = \sum_{n_1,\dotsc,n_D} U_{a_1n_1}\dotsm V_{a_Dn_D}\ T_{n_1\dotsc n_D}  \; ,\qquad 
 \bar T'_{  a_1\dots  a_D} = \sum_{n_1,\dotsc,n_D} \bar U_{  a_D n_D}\dotsm \bar V_{a_1 n_1}\ \bar T_{n_1\dots n_D}  \; .
\eea 
We will denote the indices of the complex conjugated tensor with a bar, 
and use the shorthand notation $\vec n =(n_1, \dotsc, n_D)$. We restrict from now on to $N_i =N, \; \forall i$. 

Among the invariants built out of $T$ and $\bar T$ we will deal in the sequel exclusively with 
\emph{trace invariants}. They are obtained by contracting two by two 
covariant with contravariant indices in a polynomial in the tensor entries,
\bea
\Tr (T,\bar T) = \sum \prod \delta_{n_1,\bar n_1} \;  T_{n_1\dotsc} \dots \bar T_{\bar n_1 \dots } \; ,
\eea 
where \emph{all} indices are saturated. A trace invariant has the same number of $T$ and $\bar T$. 
As $T_{n_1,\dots n_D}$ transforms as a complex vector under the action of the unitary group on only one index,
one can use the fundamental theorem of classical invariants for $U(N)$ (whose origins can be traced back to Gordan \cite{Gordan}, see
\cite{Malek} and references therein) successively for each index and conclude that {\it all} invariant polynomials in the tensor entries
write as linear combination of trace invariants.

\begin{figure}[ht]
\begin{center}
 \includegraphics[width=8cm]{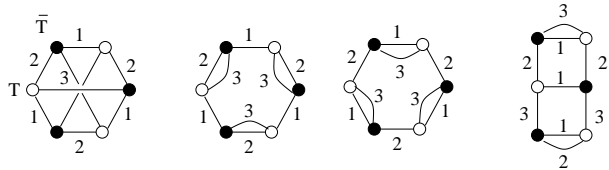}  
\caption{Graphical representation of trace invariants.}
\label{fig:tensobs}
\end{center}
\end{figure}

Trace invariants are labeled by graphs. To draw the graph associated to a trace invariant we 
represent every $T$ by a white vertex $v$ and every $\bar T$ by a black vertex $\bar v$. We promote the position of an index 
to a color: $n_1$ has color $1$, $n_2$ has color $2$ and so on. The contraction of two indices $n_i$ and $\bar n_i$ of two tensors
is represented by a line $l^i = (v,\bar v)$ connecting the corresponding two vertices. Lines inherit the color of the index, and 
always connect a black and a white vertex. Any trace invariant is thus represented by a closed {\it $D$-colored graph} $\cB$ 
\cite{coloredreview}. The graph $\cB$ is bipartite (it has black and white vertices denoted $\bar v$ and $v$, such that all lines
connect a black and a white vertex) and its lines $l^i = (v,\bar v)$ have a color $i=1,2,\dots D$. All the vertices of $\cB$ have 
coordination $D$ with all the edges incident to a given vertex having distinct colors.
Some trace invariants for rank 3 tensors are represented in figure \ref{fig:tensobs}.

The trace invariant associated to the graph $\cB$ writes as 
\bea
\Tr_{\cB}(T,\bar T ) = \sum_{\{\vec n^v,\bar{\vec{n}}^v\}_{v,\bar v \in \cV}}  \delta^{\cB}_{\{\vec{n}^v, \bar{\vec{n}}^{\bar v}\}}  \; 
\prod_{v,\bar v \in \cB} T_{\vec n^v} \bar T_{\bar {\vec n }^{\bar v} } 
\; ,\quad \text{with} \qquad \delta^{\cB}_{\{\vec{n}^v, \bar{\vec{n}}^{\bar v}\}} = \prod_{i=1}^D 
\prod_{l^i = (v,\bar v)\in \cB} \delta_{n_i^v \bar n_i^{\bar v}} \; .
\eea
where $l^i$ runs over all the lines of color $i$ of $\cB$. The product of Kronecker delta's encoding the index contractions of the 
observable associated to the graph $\cB$, $\delta^{\cB}_{\{\vec{n}^v, \bar{\vec{n}}^{\bar v}\}}$  is called the \emph{trace invariant operator 
with associated graph} $\cB$ \cite{Gurau:2011tj}. Trace invariant operators factor over the connected 
components of the graph. We denote $\Gamma_{2k}^{ ( D ) }$ the set of $D$-colored, connected graphs with $2k$ unlabeled vertices. 

Formally, a $D$ colored graph with labeled vertices is defined by an incidence matrix $\epsilon_{v\bar v}$ whose 
entries are $\{ i_1,\dots i_k \}$ if the vertices $v$ and $\bar v$ are connected by lines of colors 
$i_1, \dots, i_k$. In particular $\epsilon_{v\bar v} = \emptyset$ if $v$ and $\bar v$ are not connected by any line. An 
element $\cB \in \Gamma_{2k}^{(D)}$ is an equivalence class of incidence matrices related to one another by permutations of 
lines and columns (corresponding to relabellings of the vertices). One can write the trace invariant associated to $\cB$ 
directly in terms of the incidence matrix of any representative graph with labeled vertices 
$\prod_{v, \bar v } \prod_{i \in \epsilon_{v\bar v} } \delta_{n^v_i \bar n^{\bar v}_i} $.

The subgraphs with two colors of a $D$-colored graph $\cB$ are called \emph{faces}. We denote the number of faces of a 
graph $\cB$ by $F_{\cB}$. They will play an important role in the next section. 
The graphs with $D=3$ colors represented in figure \ref{fig:tensobs} 
posses three type of faces, given by the subgraphs with lines of colors $(1,2)$, $(1,3)$ and $(2,3)$.

To every graph $\cB$ we can associate a non negative integer, its \emph{degree} $\omega(\cB)$ \cite{GurRiv,Gur4,coloredreview}. 
The main property of the degree is that it provides a counting of the number of faces $F_{\cB}$: for a closed, connected graph 
with $D$ colors and $2p_{\cB}$ vertices the total number of faces is \cite{Gur4,coloredreview,Bonzom:2012hw}
\bea \label{eq:facesmese}
  F_{\cB} = \frac{(D-1)(D-2)}{2} p_{\cB} + (D-1) - \frac{2}{(D-2)!} \; \omega(\cB) \; .
\eea
The degree provides in higher dimensions a generalization of the genus of ribbon graphs, and indexes their
the $1/N$ expansion. It is however {\it not} a topological invariant,
but it combines topological and combinatorial information about the graph \cite{Bonzom:2011br}. 
Of course the degree can be defined for graphs
with $D+1$ colors (say $0,1,\dots D$), and the number of faces of graph with $D+1$ colors is
$F_{\cG} = \frac{D(D-1)}{2} p_{\cG} + D - \frac{2}{(D-1)!} \; \omega(\cG)$. Another important property of the degree 
(see for instance \cite{coloredreview}) is the following. Consider a closed connected graph $\cG$ with $D+1$ 
colors $0,1,\dots D$, and denote $\cB_{\rho}$ its connected subgraphs
of colors $1,\dots D$ (where $\rho=1\dots |\rho|$ labels the connected components). Then
\bea\label{eq:bound}
 \omega(\cG) \ge D \sum_{\rho=1}^{|\rho|} \omega(\cB_{\rho}) \; .
\eea

Going back to invariants one can build out of a complex tensor, we note that there exists a unique $D$-colored graph with 
two vertices, namely the graph in which all the lines connect the two vertices. We call it the $D$-dipole and its associated invariant 
is
\bea\label{eq:gaussian}
 \Tr_{\mathrm{dipole}} (T , \bar T ) = 
\sum_{\vec n,\bar{\vec{n}}}\, T_{\vec n}\, \bar T_{\bar {\vec n} }\ \Bigl[\prod_{i=1}^D \delta_{n_i \bar n_i}\Bigr] \; .
\eea 

The most general single trace invariant action for a non-symmetric tensor is 
\bea \label{eq:actiongen}
 S(T,\bar T) = t_1\, \Tr_{ \mathrm{dipole} } (T , \bar T ) + \sum_{k=2}^{\infty} \sum_{\cB\in \Gamma_{2k}^{ ( D )} } t_{\cB}\, 
\Tr_{\cB}(T,\bar T)\;,
\eea
where $t_\cB$ are the coupling constants associated to the $D$-colored graphs $\cB$ and we singled out the quadratic part corresponding to 
the $D$-dipole. 
In equation \eqref{eq:actiongen} one sometimes adds a scaling factor $N^{-\frac{2}{(D-2)!} \omega(\cB)}$ for every trace invariant
(as $\omega(\cB)\ge 0$ this suppresses some of them). Adding this extra scaling would simplify 
some formulae but does not modify anything in the sequel. We will treat in this paper the most general single 
trace rank $D$ tensor model defined by the partition function
\be 
Z(t_\cB) = \exp\bigl(-F(t_\cB)\bigr) = \int d\bar T dT\ \exp\ \Bigl(-N^{D-1} S (T,\bar  T)\Bigr) \;.
\ee

\subsection{Graph amplitudes}

\begin{figure}[ht]
\begin{center}
 \includegraphics[width=3cm]{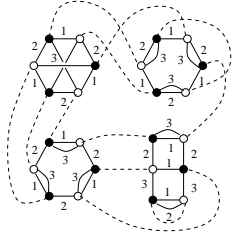}  
\caption{A Feynman graph of tensor models.}
\label{fig:tensobsgraph}
\end{center}
\end{figure}

The invariant observables are the trace invariants represented by $D$-colored graphs. The Feynman graphs 
contributing to the expectation of an observable are obtained by Taylor expanding with respect to $t_{\cB}$ and evaluating
the Gaussian integral in terms of Wick contractions. The $D$-colored graphs $\cB$ associated to the invariants 
$ \Tr_{\cB}(T,\bar T) $ in the action act as {\it effective Feynman vertices}. The effective vertices are 
connected by effective \emph{propagators} (Wick contractions, pairings of $T$'s and $\bar T$'s). 
A Wick contraction of two tensor entries $T_{a_1\dotsc a_D}$ and $\bar T_{\bar p_1 \dotsc \bar p_D}$ 
with the quadratic part \eqref{eq:gaussian} consists in replacing them by 
$ \frac{ 1}{N^{D-1} t_1} \prod_{i=1}^D \delta_{a_i \bar p_i} $. 
We will represent the Wick contractions as dashed lines labeled by the fictitious color $0$.
The dashed lines of color $0$ are thus very different from the solid lines of colors $1,2,\dots D$: they identify \emph{all} 
the indices of the two vertices (one white corresponding to $T$ and one black corresponding to $\bar T$) it connects
(recall that the lines of colors $1,2,\dots D$ identify only one index each). An example of a Feynman graph is presented in figure \ref{fig:tensobsgraph}. 

The Feynman graphs $\cG$ are therefore $(D+1)$-colored graphs. We will keep the notation $\cB$ for the $D$-colored graphs
and denote $\cG$ the $(D+1)$-colored graph.
We call the connected components with colors $1,\dots D$ of $\cG$, denoted $\cB_{\rho}$ with $\rho = 1, \dots |\rho|$, the {\it $D$-bubbles}
of $\cG$. For instance the $3$-bubbles with colors $123$ of the graph in figure \ref{fig:tensobsgraph} are the subgraphs made of
solid lines.

A graph $\cG$ has two kinds of faces: those with colors  
$i,j=1,\dotsc,D$ (which belong also to some D-bubble $\cB$), and those with colors $0,i$, for $i=1,\dotsc,D$. 
We denoted $F_{\cG}^{0i}$ the number of faces of colors $0i$ of $\cG$.  

The free energy has an expansion in closed, connected $(D+1)$-colored graphs,
\be
F(t_{\cB}) = \sum_{\cG\in \Gamma^{(D+1)}} \frac{ (-1)^{|\rho|}}{s(\cG)}\ A(\cG) \;,
\ee
where $s(\cG)$ is a symmetry factor and $|\rho|$ is the number of effective vertices (that is subgraphs
with colors $1\dotsc D$ or $D$-bubbles). The amplitudes of $\cG$ is
\be
A(\cG) = \prod_{\rho} t_{\cB_{\rho}} \; \sum_{\{\vec{n}^v,\bar{\vec{n}}^{\bar v}\}}\, \Bigl[
\prod_{\rho}  N^{D-1} \delta^{\cB_{\rho}}_{\{\vec{n}^v,\bar{\vec{n}}^{\bar v}\}}\Bigr] 
\Bigl[\prod_{l^0=(v,\bar v) \in \cG } \frac{1}{t_1 N^{D-1}} \prod_{i} \delta_{n_i^v,\bar n_{i}^{\bar v}} \Bigr]\;.
\ee
An index $n_i$ is identified along the lines of color $i$ in $\cB_{\rho}$ and along the dashed lines of color $0$. We thus obtain a free 
sum per face of colors $0i$, so that
\be
A(\cG) =  \frac{\prod_{\rho} t_{\cB_{(\rho)}}}{t_1^{ |l^0| }} 
   \ N^{(D-1)|\rho|  - (D-1) |l^0| + \sum_i |F_{\cG}^{0i}| } \;,
\ee
where $|l^0|$ denotes the total number of lines of color $0$ of $\cG$.
But $\sum_{i}F_{\cG}^{0i}= F_{\cG}- \sum_{\rho}F_{\rho}$, where $F_{\cG}$ denotes the 
total number of faces of $\cG$ and $F_{\rho}$ the number of faces of the $D$-bubble $\cB_{\rho}$.
Using \eqref{eq:facesmese} for each $\cB_{\rho}$ and for $\cG$ (taking into account that $\cG$ has $D+1$ colors)
and noting that $|l^0|=p$, with $p$ the half number of vertices of $\cG$, we obtain
\bea\label{eq:amplifini}
A(\cG) &=& \frac{\prod_{\rho} t_{\cB_{(\rho)}}}{t_1^{ p }} \ N^{D -\frac{2}{(D-1)!} \omega(\cG) + \frac{2}{(D-2)!}\sum_{\rho=1}^{|\rho|} 
\omega(\cB_{\rho})  }
\crcr
& = & \frac{\prod_{\rho} t_{\cB_{(\rho)}}}{t_1^{ p }} \ N^{D - \frac{2}{ D!} \omega(\cG) - \frac{2}{D (D-2)!}
\Bigl( \omega(\cG) -  D \sum_{\rho=1}^{|\rho|} \omega(\cB_{\rho})  \Bigr) } \;,
\eea
with $\omega(\cG)$ the degree of the graph $\cG$ and $\omega(\cB_{\rho})$ the degree of $\cB_{\rho}$.
The amplitude of a graph $\cG$ is thus at most $N^D$ and it is suppressed with the degree $\omega(\cG)$. This is the 
$1/N$ expansion for random tensor models \cite{Gur4, Gurau:2011tj, Bonzom:2012hw}. Expectation values of 
the observables have similar expansions in $1/N$ 
\bea
&& \frac{    \Big{\langle}  \Tr_{\cB_{1}}(T,\bar T) \dots \Tr_{\cB_{|\alpha|}}(T,\bar T)    \Big{\rangle}_c    }{Z}  =  
  \sum_{\cG \supset \cup_{\alpha} \cB_{\alpha} }
   N^{D - |\alpha| (D-1) -\frac{2}{ (D-1)!} \omega(\cG) + \frac{2}{(D-2)!}  
\sum_{\rho=1 }^{|\rho|}   \omega(\cB_{\rho})} \frac{\prod_{\rho\neq \alpha} t_{\cB_{\rho}}}{t_1^{ p }} \; .
\eea
where the sum runs over all $D+1$ colored graphs $\cG$ with $|\alpha|$ marked $D$ -bubbles $ \cB_{\alpha} $,
and $ \cB_{\rho} $ denotes all the $D$-bubbles of $\cG$ (hence the $\alpha$'s are some of the $\rho$'s).

\section{Schwinger Dyson equations}\label{sec:SDE}
 
The Schwinger Dyson equations of the model write in terms of two graphical operations. They encode the changes an observable $\cB$
undergoes when adding a line of color $0$. The $D$-colored graphs $\cB$ represent $D-1$ dimensional closed connected 
pseudomanifolds \cite{lost}. When seen as subgraphs of some $D+1$ colored graph $\cG$ they become chunks of a $D$ dimensional
space. The chunk associated to $\cB$ is the topological cone over the pseudomanifold \cite{Bonzom:2012hw}. The same holds for matrix 
models: the invariants $\Tr[(MM^{\dagger})^q]$ are either seen as loop observables, or as polygons when acting as effective Feynman 
vertices. The polygon represented by $\Tr[(MM^{\dagger})^q] $ when it appears as an effective vertex in a Feynman graph is the 
cone over the loop. 

The usual loop equations of matrix models encode the change of the loop observables by the addition 
of a line: a loop can either merge with another loop or it can split into two loops. In the dual triangulation
this is seen as the gluing of two segments bounding loops. Either the two segments belong to two distinct loops, and then one
obtains a larger loop, or they belong to the same loop, in which case the loop splits into two. A loop
observable $\Tr[(MM^{\dagger})^{q+1}] $ is fully characterized by its ``length'' $q$. The merging of two loops of length $p$ and $q$ 
leads to a loop with length $p+q$: the merging is just a graphical representation of the addition of integers. The SDE's of the model
are indexed by loops, hence by integers. They can be transformed into a set of constraints on the partition function $L_p Z=0$.
The latter respect the Virasoro algebra \cite{Ambjorn:1990ji,Fukuma:1990jw,Makeenko:1991ry,Dijkgraaf:1990rs}, $[L_p,L_q] = (p-q) L_{p+q}$. 
Graphically
this equation is understood as follows. Given two loops of length $p$ and $q$, in how many ways can I compose them? I can either glue the 
loop of length $q$ on the loop of length $p$ (and one has $p+1$ choices for where to insert the second loop), or I can 
glue the loop of length $p$ on the loop of length $q$ (and one has $q+1$ choices for where to perform this insertion)
In all cases I obtain a loop of length $p+q$, associated to the operator $L_{p+q}$.

A similar picture holds, with the appropriate generalizations, in higher dimensions.

\subsection{Graph operations}

{\it The colored gluing of two graphs}. Consider two $D$ colored graphs $\cB_1$ and $\cB_2$ and chose two vertices
$v_1\in \cB_1$ and $\bar v_2 \in \cB_2$.  We define the colored gluing of graphs at $v_1$ and $\bar v_2$ as the graph
$  \cB_1 \star_{(v_1,\bar v_2)} \cB_2$ obtained by deleting the vertices $v_1$ and $\bar v_2$ and joining all the lines
touching them pairwise respecting the colorings. 

\begin{figure}[ht]
\begin{center}
 \includegraphics[width=8cm]{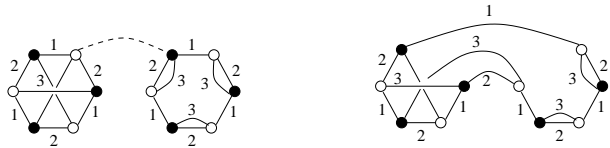}  
\caption{Graphical representation of the colored gluing of two graphs.}
\label{fig:composition}
\end{center}
\end{figure}

This operation can be performed in two steps. Consider the trace invariants associated to $\cB_1$ and $\cB_2$ and connect 
the two vertices $v_1$ and $\bar v_2$ by a dashed line of color $0$. Forgetting for a second the scalings with $N$,
this line will identify all the indices of the tensor associated to $v_1$ with the ones of the tensor associated to $\bar v_2$
\bea
\sum_{ n_i^{v_1} , \bar n_i^{\bar v_2} }  \delta^{\cB_1}_{ \{ \vec n^v,\bar {\vec n}^{\bar v} \} }  \;
  \Bigl( \prod_i \delta_{n_i^{v_1} \bar n_i^{\bar v_2}} \Bigr) \;
\delta^{\cB_2}_{ \{ \vec n^v,\bar {\vec n}^{\bar v} \} } &=&
\sum_{ n_i^{v_1} , \bar n_i^{\bar v_2} } \Bigl( \prod_{i=1}^D 
\prod_{l^i = (v,\bar v)\in \cB_1} \delta_{n_i^v \bar n_i^{\bar v}}    \Bigr) \Bigl( \prod_i \delta_{n_i^{v_1} \bar n_i^{\bar v_2}} \Bigr)
\Bigl( \prod_{i=1}^D  \prod_{l^i = (v,\bar v)\in \cB_2} \delta_{n_i^v \bar n_i^{\bar v}}    \Bigr) \crcr
&=& \delta^{   \cB_1 \star_{(v_1,\bar v_2)} \cB_2   }_{  \{ \vec n^v,\bar {\vec n}^{\bar v} \}  } \; ,
\eea 
as $v_1$ is a vertex in $\cB_1$ and $\bar v_2$ is a vertex in $\cB_2$. The gluing is represented graphically
in figure \ref{fig:composition}.

Of course the gluing preserves the colorability, hence $  \cB_1 \star_{(v_1,\bar v_2)} \cB_2  $ is a connected 
$D$-colored graph with colors $1,\dots D$.
At the level of the incidence matrix, one builds from $\epsilon^{\cB_1}_{v\bar v}$ and $\epsilon^{\cB_2}_{v\bar v} $
the incidence matrix
\bea
  v\neq v_1 ,  \bar v\neq \bar v_2  \qquad \epsilon^{\cB_1 \star_{(v_1,\bar v_2)} \cB_2  }_{v\bar v} = \begin{cases}
                                                               \epsilon^{\cB_1}_{v\bar v} \text{ if } v,\bar v \in \cB_1  \\
                                                               \epsilon^{\cB_2}_{v\bar v} \text{ if } v,\bar v \in \cB_2  \\
                                                               \epsilon^{\cB_1}_{v_1\bar v} \cap \epsilon^{\cB_2}_{v\bar v_2} \text{ if } v \in \cB_2, 
                                                                             \bar v \in \cB_1  
                                                             \end{cases} \;
= \epsilon^{\cB_1}_{v\bar v} \cup  \epsilon^{\cB_2}_{v\bar v} \cup 
  \Bigl(  \epsilon^{\cB_1}_{v_1\bar v} \cap \epsilon^{\cB_2}_{v\bar v_2}   \Bigr) .
\eea

{\it The colored contraction of a graph.} The second operation is similar to the gluing, but it pertains to an unique graph.
Let $\cB_1$ be a $D$ colored graph and select two vertices $v_1,\bar v_1 \in \cB_1$. The contraction of $\cB_1$ with the pair of vertices,
denoted $\cB_1/(v_1,\bar v_1)$ is the graph obtained from $\cB_1$ by deleting the vertices $v_1$ and $\bar v_1$ and reconnecting the 
lines touching them pairwise respecting the colorings. 

\begin{figure}[ht]
\begin{center}
 \includegraphics[width=4cm]{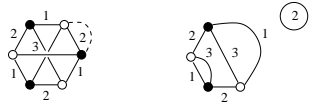}  
\caption{Graphical representation of the contraction of a graph.}
\label{fig:reduction}
\end{center}
\end{figure}

Again this operation can be performed in two steps. Consider the trace invariant associated to $\cB_1$ and connect 
the two vertices $v_1$ and $\bar v_1$ by a dashed line of color $0$. Again 
this line will identify all the indices of the tensor associated to $v_1$ with the ones of the tensor associated to $\bar v_1$
\bea
\sum_{ n_i^{v_1} , \bar n_i^{\bar v_1} }  \delta^{\cB_1}_{ \{ \vec n^v,\bar {\vec n}^{\bar v} \} }  \;
  \Bigl( \prod_i \delta_{n_i^{v_1} \bar n_i^{\bar v_1}} \Bigr) \;
&=&
\sum_{ n_i^{v_1} , \bar n_i^{\bar v_1} } \Bigl( \prod_{i=1}^D 
\prod_{l^i = (v,\bar v)\in \cB_1} \delta_{n_i^v \bar n_i^{\bar v}}    \Bigr) \Bigl( \prod_i \delta_{n_i^{v_1} \bar n_i^{\bar v_1}} \Bigr) \crcr
&=& \delta^{   \cB/(v_1,\bar v_1) }_{  \{ \vec n^v,\bar {\vec n}^{\bar v} \}  } \; ,
\eea 
as both $v_1$ and $\bar v_1$ are vertices in $\cB_1$. The contraction is represented graphically
in figure \ref{fig:reduction}.

The contraction preserves the colorability. Note that the graph $ \cB/(v_1,\bar v_1) $ can potentially 
be disconnected (see figure \ref{fig:reduction}). We will denote its connected components 
$ \bigl[  \cB/(v_1,\bar v_1) \bigr]_{\rho}$. Moreover, some of these connected components can consist in an 
unique line (as it is the case in figure  \ref{fig:reduction}). In this case they are not strictu sensu 
$D$ colored graphs, but consist in exactly one line (with some color) which closes onto itself. This happens for every 
line which connects directly $v_1$ and $\bar v_1$ in $\cB$. Such a line brings a factor $N$.
At the level of the incidence matrix, one builds from $\epsilon^{\cB_1}_{v\bar v}$ 
the incidence matrix
\bea
  v\neq v_1 ,  \bar v\neq \bar v_1  \qquad \epsilon^{\cB_1/ (v_1,\bar v_1)  }_{v\bar v} = 
             \epsilon^{\cB_1}_{v\bar v}  \cup \Bigl( 
     \epsilon^{\cB_1}_{v_1\bar v} \cap \epsilon^{\cB_1}_{v\bar v_1} \Bigr) \; .
\eea
and the connected components with no vertices of course do not have an incidence matrix.

These two operations encode the changing of an observable when adding a line of color $0$. If the 
line of color $0$ in $\cG$ is a tree line (joins $\cB_1$ with some $\cB_2$), then the observables $\cB_1$ and 
$\cB_2$ are glued.
If on the other hand $l^0$ is a loop line (starts and ends on the same $\cB_1$), then the observable 
is contracted. 

\begin{figure}[ht]
\begin{center}
 \includegraphics[width=9cm]{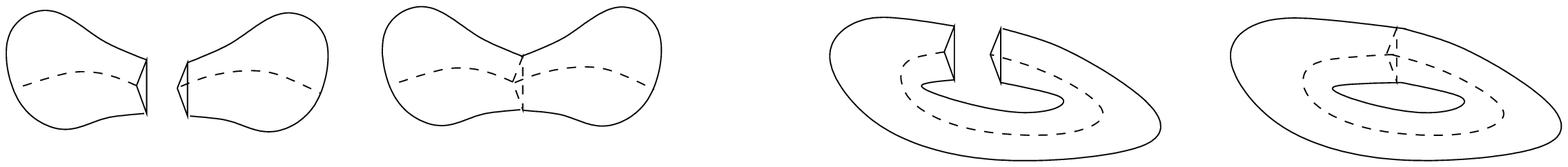}  
\caption{Gluing and contraction of surfaces for tensors of rank $D=3$.}
\label{fig:gluered}
\end{center}
\end{figure}

Geometrically these two operations have the following interpretation. Set for now $D=3$. Then $\cB_1$ and 
$\cB_2$ represent surfaces. The surface associated to $\cB_1$ is obtained by associating a triangle
with edges colored $1$, $2$ and $3$ to each vertex of $\cB_1$. This induces a coloring of the points (vertices) 
of the triangle by pairs of colors: $12$ is the point common to the edges $1$ and $2$ and so on. A line in $\cB_1$ represents
the unique gluing of the two triangles corresponding to its end vertices which respects {\it all} the 
colorings (those of the edges {\it and} those of the points,
see \cite{Bonzom:2012hw} for more details). The gluing $ \cB_1 \star_{(v_1,\bar v_2)} \cB_2  $ comes to choosing a triangle
(corresponding to $v_1$) on $\cB_1$, a triangle (corresponding to $\bar v_2$) on $\cB_2$ and gluing the two surfaces along
the triangles. This is represented in figure \ref{fig:gluered} on the left, where we depicted the simplest case
of the gluing of two planar surfaces. The contraction is essentially the same thing,
just that this time the two triangles belong to the same surface, as represented in figure \ref{fig:gluered} on the right.

Note that the topology of the surfaces is changed under these moves (in the example of figure \ref{fig:gluered} on the right 
a planar surfaces becomes a
genus one surface). This should come as no surprise as the same happens for matrix models: the contraction of a loop leads to two loops.
The gluing is just a graphical encoding of surgery on the surfaces. The contraction has a more involved topological interpretation, and
it can lead to an increase of the genus (if the two triangles contracted do not share anything), a splitting of the surface
into connected components
(if the two triangles contracted share at least two vertices but no edges) or no change at all (if the two triangles contracted share edges).

The most important feature of the gluing and the contraction is the following. Consider two observables 
$\cB_1$ and $\cB_2$ joined by two lines of color $0$, $(v_1, \bar v_2)$ and $(v_2,\bar v_1)$  with $v_1,\bar v_1 \in \cB_1$
and $v_2,\bar v_2 \in \cB_2$. The resulting observable can be obtained in two ways, either by gluing along 
$(v_1, \bar v_2)$ and contracting with respect to $(v_2,\bar v_1)$ or the reverse. As the end result is unique, we have
\bea\label{eq:doubleline}
   v_1,\bar v_1 \in \cB_1 \;, \;  v_2,\bar v_2 \in \cB_2 \Rightarrow 
  \bigl[  \cB_1 \star_{ (   v_1, \bar v_2 ) }\cB_2 \bigr] / (v_2,\bar v_1)   = 
   \bigl[  \cB_2 \star_{   (v_2,\bar v_1) }\cB_1 \bigr] /( v_1, \bar v_2 )  \; .
\eea

In $D=2$ dimensions the observables are just bi colored cycles. The gluing of two cycles of lengths $p$ and $q$ 
always leads to a cycle of length $p+q$ thus the gluing reduces to the addition of the lengths (hence it is associative).
The colored gluing of graphs is associative (provided one tracks the vertices at which 
it is made) only if both expressions $ [\cB \star_{(v,\bar v_1)} \cB_1] \star_{(v', \bar v_2)}\cB_2 $ 
and $\cB \star_{(v,\bar v_1)} [\cB_1 \star_{(v', \bar v_2)}\cB_2 ]$ are defined, that is if $v'\in \cB_1$. 
Note however that if $v, v' \in \cB$, while 
$[\cB \star_{(v,\bar v_1)} ]\star_{(v', \bar v_2)}\cB_2$ is defined, $\cB \star_{(v,\bar v_1)} [\cB_1 \star_{(v', \bar v_2)}\cB_2 ]$
is {\it not}. The gluing is the appropriate generalization of this addition to the  $D$-colored graphs representing the 
observables of tensor models. In the spirit of the matrix model nomenclature, one should call the trace invariants 
observables ``bubble observables'', and the SDE's we derive in the next section ``bubble equations''.

\subsection{Schwinger Dyson equations and the algebra of constraints}

We now derive the SDE's of tensor models at all orders in $N$. We subsequently
translate them into constraints satisfied by the partition function. The constraints 
form a Lie algebra, generalizing to all orders in $N$ the $D$-ary tree algebra identified in \cite{Gurau:2011tj}.

Consider a $D$-colored graph $\cB_1$. We chose a vertex $\bar v_1 \in \cB$
(and mark it). For any $\cB_1$ and $\bar v_1\in \cB_1$ the following trivial identity holds
\bea
&& \sum_{\vec p; n,\bar n} \int [dT d\bar T] \; \frac{\delta}{\delta T_{\vec p}} \Bigl( \delta_{ \bar {\vec n}_{\bar v_1} \vec p} 
\prod_{v\in \cB} T_{ \vec  n_{ v}}   \prod_{ \bar v \in \cB_1 \; \bar v \neq \bar v_1} 
   {\bar T}_{ \bar {\vec n}_{ v}}  \; 
   \delta^{\cB_1}_{n\bar n}  \; e^{-N^{D-1} S }\Bigr) =0  \Rightarrow \crcr
&& \Big \langle  \sum_{v_1 \in \cB_1 } \Tr_{\cB_1{/(v_1, \bar  v_1 )} } (T,\bar T)   \Big\rangle - N^{D-1} \sum_{\cB}  t_{\cB }  
\sum_{ v \in \cB}  \Big \langle  \sum \Tr_{\cB\star_{(v,\bar v_1)} \cB_1  } (T,\bar T)   \Big\rangle =0 
\; .
\eea
 
We denote the $|\rho|$ connected components of $\cB_1 / (   v_1, \bar  v_1  )$ by
$ \bigl[ \cB_1 / (   v_1, \bar  v_1  )  \bigr]_{\rho}$, with $1 \le |\rho| \le D$. If one of these connected 
components consist in a single line the associated trace is $N$. Thus 
\bea
  \Tr_{\cB_1 / (v_1, \bar  v_1 )} (T,\bar T) = \prod_{ \rho=1}^{|\rho|}  
   \Tr_{ \bigl[ \cB_1 / (v_1, \bar  v_1 ) \bigr]_{\rho} } (T,\bar T) \; .
\eea
The SDE's translates into a set of differential operators acting on $Z$ indexed by the observable $\cB_1$ and the marked vertex $\bar v_1$
\bea
 && \cL_{(\cB_1,\bar v_1)}Z = 0 \; ,\crcr
 && \cL_{(\cB_1,\bar v_1)} = \sum_{   v_1 \in \cB_1 }\prod_{ \rho =1 }^{|\rho|}  \Bigl ( -  \frac{1}{N^{D-1  } }
  \frac{\partial } {  \partial t_{   \bigl[ \cB_1 / (v_1, \bar  v_1 ) \bigr]_{\rho}   } } \Bigr)  
  + \sum_{\cB}  t_{\cB }    
\sum_{ v \in \cB} \frac{\partial}{ \partial t_{  \cB\star_{(v,\bar v_1)} \cB_1      }} \; ,
\eea
and by convention the partial derivative is $-N^D$ if $ \bigl[ \cB_1 / (v_1, \bar  v_1 ) \bigr]_{\rho}  $ 
is formed by an unique line. The natural domain of the differential operators 
$\cL_{(\cB_1,\bar v_1)} $ is the set of invariant functions depending on the coupling constants $t_{\cB}$.
The constraints form a Lie algebra

\begin{theorem}
 The commutator of two differential operators $\cL_{(\cB_1,\bar v_1)}  $ and $  \cL_{(\cB_1,\bar v_1)}  $ is
\bea
 \Bigl[  \cL_{(\cB_1,\bar v_1)}   , \cL_{(\cB_2,\bar v_2)}     \Bigr] =
 \sum_{v\in \cB_1 }\cL_{ \bigl(  \cB_1  \star_{ ( v,\bar v_2 ) } \cB_2 , \bar v_1   \bigr) }
  - \sum_{v\in \cB_2} \cL_{   \bigl(  \cB_2  \star_{ ( v,\bar v_1 ) } \cB_1 , \bar v_2   \bigr)       } \; .
\eea
\end{theorem}

{\bf Proof:} The proof is a straightforward computation. We start from the commutator
\bea\label{eq:comutator}
&&  \Bigl[  \cL_{(\cB_1,\bar v_1)}   , \cL_{(\cB_2,\bar v_2)}     \Bigr] \crcr
&&= \Bigl[   \sum_{  v_1 \in \cB_1 }\prod_{ \rho =1 }^{|\rho|}  \Bigl ( -  \frac{1}{N^{D-1  } }
  \frac{\partial } {  \partial t_{   \bigl[ \cB_1/ (v_1, \bar  v_1 ) \bigr]_{\rho}   } } \Bigr) ,
  \sum_{\cB}  t_{\cB }    
\sum_{ v \in \cB} \frac{\partial}{ \partial t_{  \cB\star_{(v,\bar v_2)} \cB_2      }} \Bigr] 
\crcr
&&- \Bigl[ \sum_{\cB}  t_{\cB }    
\sum_{ v \in \cB}   \frac{\partial}{ \partial t_{  \cB\star_{(v,\bar v_1)} \cB_1      }}  , 
\sum_{  v_2 \in \cB_2 }\prod_{ \rho =1 }^{|\rho|}  \Bigl ( -  \frac{1}{N^{D-1  } }
  \frac{\partial } {  \partial t_{   \bigl[ \cB_2/ (v_2, \bar  v_2 ) \bigr]_{\rho}   } } \Bigr)  
 \Bigr] \crcr
&&+ \Bigl[\sum_{\cB}  t_{\cB }    
\sum_{ v \in \cB} \frac{\partial}{ \partial t_{  \cB\star_{(v,\bar v_1)} \cB_1      }} ,  
\sum_{\cB'}  t_{\cB' }    
\sum_{ v' \in \cB'}  \frac{\partial}{ \partial t_{  \cB'\star_{(v',\bar v_2)} \cB_2      }}
 \Bigr] \; .
\eea
The first line evaluates to
\bea
&&  \sum_{ v_1 \in \cB_1 } \sum_{\mu=1}^{|\rho|}  \prod_{ \rho =1, \rho \neq \mu }^{|\rho|}  
   \Bigl ( -  \frac{1}{N^{D-1  } }
  \frac{\partial } {  \partial t_{   \bigl[ \cB_1/ (v_1, \bar  v_1 ) \bigr]_{\rho}   } } \Bigr)
   \sum_{ v \in \bigl[ \cB_1/ (v_1, \bar  v_1 ) \bigr]_{\mu}   }   
 \Bigl ( -  \frac{1}{N^{D-1  } } \Bigr)
 \frac{\partial}{ \partial t_{    \bigl[ \cB_1/ (v_1, \bar  v_1 ) \bigr]_{\mu}   \star_{(v,\bar v_2)} \cB_2      }} \; .
\eea

Consider the bubble $  \cB_1 \star_{(v, \bar v_2)} \cB_2 $. When reducing with respect to $   (v_1, \bar  v_1 )$ 
with $ v_1$ in $\cB_1$ it will disconnect into several connected components 
$  \Bigr{\{} \Bigl[ \cB_1 \star_{(v, \bar v_2)} \cB_2  \Bigr] / (v_1,\bar v_1) \Bigl{\}}_{\rho} $. All save one (the one to which $v$ belongs)
coincide with the connected components $ \bigl[ \cB_1/ (v_1, \bar  v_1 ) \bigr]_{\rho}     $.
The special one is $  \Bigr{\{} \Bigl[ \cB_1 \star_{(v, \bar v_2)} \cB_2  \Bigr] / (v_1,\bar v_1) \Bigl{\}}_{\mu}
= \bigl[ \cB_1/ (v_1, \bar  v_1 ) \bigr]_{\mu}   \star_{(v,\bar v_2)} \cB_2   
 $. Also, $  \sum_{\mu=1}^{|\rho|}   \sum_{ v \in \bigl[ \cB_1/ (v_1, \bar  v_1 ) \bigr]_{\mu}   }  
=\sum_{v\in \cB_1, v\neq v_1} $ hence the first line is
\bea
   \sum_{ v_1 \in \cB_1 } \sum_{v\in \cB_1, v\neq v_1}  \prod_{ \rho =1  }^{|\rho|}  
     \Bigl ( -  \frac{1}{N^{D-1  } }
  \frac{\partial } {  \partial t_{ \Bigr{\{} \Bigl[ \cB_1 \star_{(v, \bar v_2)} \cB_2  \Bigr] / (v_1,\bar v_1) \Bigl{\}}_{\rho}    } } \Bigr) \; ,
\eea
and exchanging the sums over $v$ and $v_1$ it becomes
\bea
 \sum_{v\in \cB_1} \sum_{v_1\in\cB_1, v_1\neq v} && \;
 \prod_{ \rho =1  }^{|\rho|}  
     \Bigl ( -  \frac{1}{N^{D-1  } }
  \frac{\partial } {  \partial t_{ \Bigr{\{} \Bigl[ \cB_1 \star_{(v, \bar v_2)} \cB_2  \Bigr] / (v_1,\bar v_1) \Bigl{\}}_{\rho}    } } \Bigr)
\crcr
=&& \sum_{v\in \cB_1} \sum_{v' \in  \cB_1 \star_{(v, \bar v_2)} \cB_2  } 
\prod_{ \rho =1  }^{|\rho|}  
     \Bigl ( -  \frac{1}{N^{D-1  } }
  \frac{\partial } {  \partial t_{ \Bigr{\{} \Bigl[ \cB_1 \star_{(v, \bar v_2)} \cB_2  \Bigr] / (v',\bar v_1) \Bigl{\}}_{\rho}    } } \Bigr) \crcr
&&- \sum_{v\in \cB_1} \sum_{v' \in\cB_2}
\prod_{ \rho =1  }^{|\rho|}  
     \Bigl ( -  \frac{1}{N^{D-1  } }
  \frac{\partial } {  \partial t_{ \Bigr{\{} \Bigl[ \cB_1 \star_{(v, \bar v_2)} \cB_2  \Bigr] / (v',\bar v_1) \Bigl{\}}_{\rho}    } } \Bigr)
\eea
where we relabeled $v_1$ by $v'$ and we added and subtracted the terms with $ v' \in\cB_2 $.
Recall that by equation \eqref{eq:doubleline}, if $v,\bar v_1 \in \cB_1$ and $v',\bar v_2 \in \cB_2$ then 
\bea
 \Bigl[ \cB_1 \star_{(v, \bar v_2)} \cB_2  \Bigr] / (v' ,\bar v_1)  = 
  \Bigl[  \cB_2 \star_{  (v' ,\bar v_1)     } \cB_1 \Bigr] /  (v, \bar v_2)  \; ,
\eea
hence the first two lines in eq. \eqref{eq:comutator} yield
\bea \label{eq:firstlines}
&& \sum_{v\in \cB_1} \sum_{v' \in  \cB_1 \star_{(v, \bar v_2)} \cB_2  } 
\prod_{ \rho =1  }^{|\rho|}  
     \Bigl ( -  \frac{1}{N^{D-1  } }
  \frac{\partial } {  \partial t_{ \Bigr{\{} \Bigl[ \cB_1 \star_{(v, \bar v_2)} \cB_2  \Bigr] / (v',\bar v_1) \Bigl{\}}_{\rho}    } } \Bigr) \crcr
&&  - \sum_{v\in \cB_2} \sum_{v' \in  \cB_2 \star_{(v, \bar v_1)} \cB_1  } 
\prod_{ \rho =1  }^{|\rho|}  
     \Bigl ( -  \frac{1}{N^{D-1  } }
  \frac{\partial } {  \partial t_{ \Bigr{\{} \Bigl[ \cB_2 \star_{(v, \bar v_1)} \cB_1  \Bigr] / (v',\bar v_2) \Bigl{\}}_{\rho}    } } \Bigr) \; .
\eea

The third line in equation \eqref{eq:comutator} writes 
\bea\label{eq:restlines}
&&  \sum_{\cB}  t_{\cB }    
\sum_{ v \in \cB} \sum_{v' \in  \cB\star_{(v,\bar v_1)} \cB_1   }
 \frac{\partial}{\partial t_{ \Bigl[   \cB\star_{(v,\bar v_1)} \cB_1    \Bigr] \star_{ ( v',\bar v_2 ) } \cB_2   } } - ( 1\leftrightarrow 2 ) \; .
\eea
We separate the terms with $ v'\in \cB_1$ from the terms with $v' \in \cB$ to get 
\bea
&& =  \sum_{\cB}  t_{\cB }    
\sum_{ v \in \cB} \sum_{v' \in \cB_1   }
 \frac{\partial}{\partial t_{ \Bigl[   \cB\star_{(v,\bar v_1)} \cB_1    \Bigr] \star_{ ( v',\bar v_2 ) } \cB_2   } }
 + \sum_{\cB}  t_{\cB }    
\sum_{ v \in \cB} \sum_{v' \in  \cB, v'\neq v   }
 \frac{\partial}{\partial t_{ \Bigl[   \cB\star_{(v,\bar v_1)} \cB_1    \Bigr] \star_{ ( v',\bar v_2 ) } \cB_2   } } 
 - ( 1\leftrightarrow 2 ) \crcr
&& = \sum_{v'\in \cB_1}\sum_{\cB}  t_{\cB } \sum_{ v \in \cB} 
  \frac{\partial}{ \partial t_{   \cB\star_{(v,\bar v_1)} \Bigl[ \cB_1  \star_{ ( v',\bar v_2 ) } \cB_2      \Bigr]    }} - ( 1\leftrightarrow 2 )
\eea

Adding equations \eqref{eq:firstlines} with \eqref{eq:restlines} we obtain, relabeling some dummy indices
\bea 
&&  \Bigl[  \cL_{(\cB_1,\bar v_1)}   , \cL_{(\cB_2,\bar v_2)}     \Bigr]  = \\
&&  \sum_{w\in \cB_1} \Bigl{\{} \sum_{v' \in  \cB_1 \star_{(w, \bar v_2)} \cB_2  } 
\prod_{ \rho =1  }^{|\rho|}  
     \Bigl ( -  \frac{1}{N^{D-1  } }
  \frac{\partial } {  \partial t_{ \Bigr{\{} \Bigl[ \cB_1 \star_{(w, \bar v_2)} \cB_2  \Bigr] / (v',\bar v_1) \Bigl{\}}_{\rho}    } } \Bigr)   
 - \sum_{\cB}  t_{\cB } \sum_{ v \in \cB} 
  \frac{\partial}{ \partial t_{   \cB\star_{(v,\bar v_1)} \Bigl[ \cB_1  \star_{ ( w,\bar v_2 ) } \cB_2      \Bigr]    }}
 \Bigr{\}}- ( 1\leftrightarrow 2 ) \; ,\nonumber
\eea
hence
\bea
  \Bigl[  \cL_{(\cB_1,\bar v_1)}   , \cL_{(\cB_2,\bar v_2)}     \Bigr] =
 \sum_{w\in \cB_1 }\cL_{ \bigl(  \cB_1  \star_{ ( w,\bar v_2 ) } \cB_2 , \bar v_1   \bigr) }
  - \sum_{w\in \cB_2} \cL_{   \bigl(  \cB_2  \star_{ ( w,\bar v_1 ) } \cB_1 , \bar v_2   \bigr)       } \; .
\eea

\qed

This Lie algebra admits a closed Lie subalgebra. The leading order observables, the melons \cite{Bonzom:2011zz}, are indexed by
colored rooted $D$-ary trees $\cT$. It is easy to check that gluing of the observables $\cB_1\star_{(v_1,\bar v_2)}\cB_2$ reproduces 
the gluing of their associated trees $\cT_1 \star_V \cT_2$, as defined in \cite{Gurau:2011tj}. The melonic observables are closed under this 
composition (as the gluing of trees leads to trees), hence the algebra indexed by $D$-ary trees identified in \cite{Gurau:2011tj}
is a Lie subalgebra of the full constraints algebra. 

\section{Conclusion}

We have derived in this paper the SDE's of tensor models for a generic complex tensor at all orders in $1/N$.
They translate into a Lie algebra of constraints obeyed by the partition function. The algebra is indexed by 
colored graphs and generalizes to all orders in $1/N$ the algebra indexed by $D$-ary trees of the 
leading order observables. The 
algebra indexed by $D$-ary trees closes a Lie subalgebra of the full constraints algebra.

The study of this algebra of constraints, primarily of its central extensions and unitary 
representations, is a prerequisite for the full classification of the continuum limits of tensor models. 
The study of its representations would benefit from identifying various Lie subalgebras, and studying
the their representations first. We already posses a candidate, the leading order algebra 
indexed by $D$-ary trees. Other, simpler subalgebras can readily be identified: for instance the Virasoro algebra itself 
is a subalgebra of the full constraints algebra \cite{doubletens} (in fact, following the results of 
\cite{doubletens}, one can identify several distinct copies of the Virasoro algebra as subalgebras of the 
full constraints algebra). The continuum SDE's should be understood in some appropriate double scaling 
limit and the continuum operators should be identified \cite{Fukuma:1990jw,Dijkgraaf:1990rs}.
Other aspects of the emergent continuous geometry like its effective spectral and Hausdorff dimensions
must also be analyzed. Analytic control of the continuum limit is a prerequisite in order to use the random 
tensor models to investigate conformal field theories, statistical models in random geometry
and quantum gravity in arbitrary dimensions.

\section*{Acknowledgements}
Research at Perimeter Institute is supported by the Government of Canada through Industry
Canada and by the Province of Ontario through the Ministry of Research and Innovation.



\end{document}